\documentclass[11pt,twoside]{article}
\usepackage{asp2014}

\aspSuppressVolSlug
\resetcounters

\begin{document}

\title{Reaching Higher Densities for Laboratory White Dwarf
  Photospheres to Measure Spectroscopic Line Profiles}
\author{Ross~E.~Falcon$^1$, J.~E.~Bailey$^1$, T.~A.~Gomez$^{1,2}$, M.~Schaeuble$^2$, T.~Nagayama$^1$, M.~H.~Montgomery$^2$, D.~E.~Winget$^2$, and G.~A.~Rochau$^1$
\affil{$^1$Sandia National Laboratories, Albuquerque, NM 87185-1196,
  USA; \email{refalco@sandia.gov}}
\affil{$^2$Department of Astronomy and McDonald Observatory, University of Texas, Austin, TX 78712, USA}}

\paperauthor{R.~E.~Falcon}{refalco@sandia.gov}{orcid.org/0000-0003-2132-4795}{Sandia National Laboratories}{}{Albuquerque}{NM}{87185-1196}{USA}
\paperauthor{J.~E.~Bailey}{jebaile@sandia.gov}{}{Sandia National
  Laboratories}{}{Albuquerque}{NM}{87185-1196}{USA}
\paperauthor{T.~A.~Gomez}{gomezt@astro.as.utexas.edu}{}{University of
  Texas}{Department of Astronomy and McDonald
  Observatory}{Austin}{TX}{78712}{USA}
\paperauthor{M.~Schaeuble}{}{}{University of Texas}{Department of
  Astronomy and McDonald Observatory}{Austin}{TX}{78712}{USA}
\paperauthor{T.~Nagayama}{tnnagay@sandia.gov}{}{Sandia National
  Laboratories}{}{Albuquerque}{NM}{87185-1196}{USA}
\paperauthor{M.~H.~Montgomery}{mikemon@astro.as.utexas.edu}{}{University of Texas}{Department of Astronomy and McDonald Observatory}{Austin}{TX}{78712}{USA}
\paperauthor{D.~E.~Winget}{dew@astro.as.utexas.edu}{}{University of
  Texas}{Department of Astronomy and McDonald
  Observatory}{Austin}{TX}{78712}{USA}
\paperauthor{G.~A.~Rochau}{garochau@sandia.gov}{}{Sandia National
  Laboratories}{}{Albuquerque}{NM}{87185-1196}{USA}

\begin{abstract}
As part of our laboratory investigation of the theoretical line
profiles used in white dwarf atmosphere models, we extend the
electron-density ($n_{\rm e}$) range measured by our experiments to
higher densities (up to $n_{\rm
  e}\sim80\times10^{16}$~cm$^{-3}$). Whereas inferred parameters using
the hydrogen-$\beta$ spectral line agree among different line-shape
models for $n_{\rm e}\lesssim30\times10^{16}$~cm$^{-3}$, we now see
divergence between models. These are densities beyond the range
previously benchmarked in the laboratory, meaning theoretical profiles
in this regime have not been fully validated. Experimentally exploring
these higher densities enables us to test and constrain different
line-profile models, as the differences in their relative H-Balmer
line {\it shapes} are more pronounced at such conditions. These
experiments also aid in our study of occupation probabilities because
we can measure these from relative line {\it strengths}.
\end{abstract}

\section{Introduction}

Theoretical line profiles are a critical ingredient of white dwarf
(WD) atmosphere models
\citep[e.g.,][]{Koester79,Bergeron92b,Koester10}. A modification to
the hydrogen line profiles by \citet{Tremblay09} resulted in
significant systematic changes to the inferred WD atmospheric
parameters (i.e., effective temperature, $T_{\rm e}$, and surface
gravity, log\,$g$) from \citet{Liebert05}. These H line profiles have
since become the standard in the community and in the comprehensive
analysis of thousands of WDs \citep[e.g.,][]{Tremblay11,Girven11,Gianninas11,Kleinman13,Limoges15,Guo15}.

Though this {\it spectroscopic} method is powerful, precise, and
widely used, its results do not agree with mass determinations using
gravitational redshifts \citep{Barstow05,Falcon10} nor inferred
atmospheric parameters using photometry
\citep{Genest-Beaulieu14}. For this latter example, H line profiles
are specifically a suspect for the disagreement.

We thus experimentally investigate the spectroscopic method by
targeting the theoretical line profiles used in WD atmosphere
models. We have performed laboratory experiments at the {\it Z}
Pulsed Power Facility
\citep[e.g.,][]{McDaniel02,Matzen05,Rose10,Savage11} at Sandia
National Laboratories to measure the spectral line profiles present
in the high-density ($n_{\rm e}$) plasmas of WD photospheres
\citep{Falcon10b,Falcon13,Montgomery15,Schaeuble16}. Having achieved
higher densities in the laboratory than previously explored in this
way---while measuring multiple spectral lines simultaneously---we now
extend our measurements to plasmas at even higher $n_{\rm e}$. This
allows us to better discriminate amongst theoretical line profiles,
since relative line {\it shapes} (i.e., among Balmer lines) differ
between calculations with increasing principal quantum number
\citep{Tremblay09} and with increasing $n_{\rm e}$. We can also
uniquely investigate occupation probabilities \citep{Hummer88} by
measuring relative line {\it strengths}.

\section{Experiment}

To reach higher electron densities, we adjust the
gas-fill pressure of our gas cell. We also spectroscopically observe the
plasma generated inside our cell \citep{Rochau14} along a line of
sight that is closer (5~mm instead of our standard 10~mm) to the gold
wall from which the photoionizing radiation emerges \citep[see][]{Falcon15c}.

\begin{figure}[!h]
\begin{center}
  \includegraphics[width=\columnwidth]{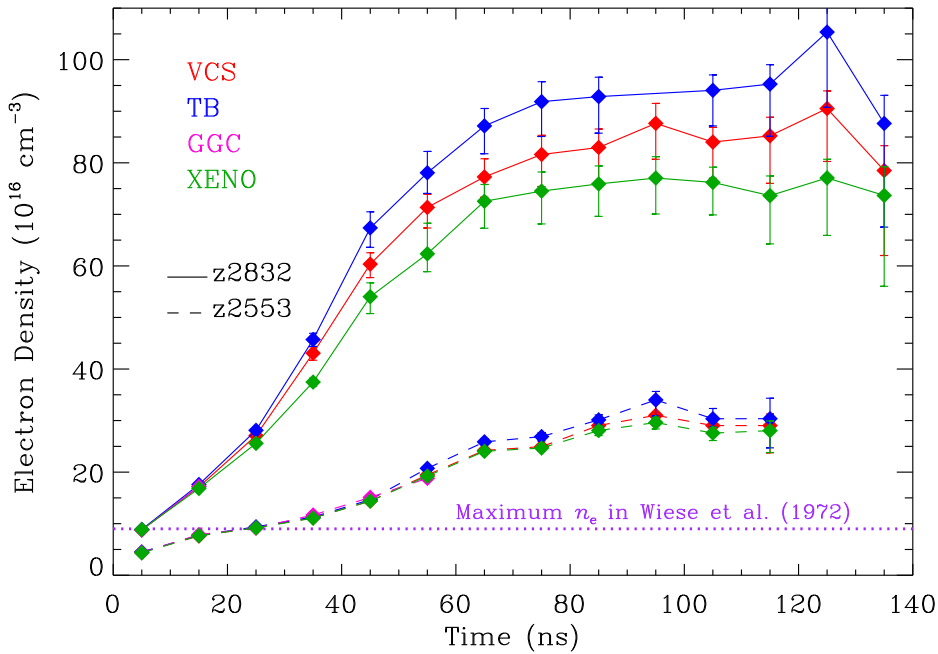}
  \caption{Electron density, $n_{\rm e}$, as a function of time throughout our
    experiments z2553 and z2832. We infer $n_{\rm e}$ using different
    theoretical line-profile calculations.}
  \label{ne}
\end{center}
\end{figure}

Figure~\ref{ne} shows our inferred $n_{\rm e}$ as a function of time
throughout two experiments. The onset of backlighting continuum
emission that allows us to measure absorption spectra of our plasma
\citep{Falcon13} occurs at 0~ns. We include data from experiment z2553
\citep{Falcon15c}, from which we infer consistent values by fitting our
measured H$\beta$ spectral line while using different theoretical
line-profile calculations. We now report data from experiment z2832,
whose $n_{\rm e}$ increases beyond that of z2553 by approximately a
factor of three. At these higher electron densities ($n_{\rm
  e}\gtrsim30\times10^{16}$~cm$^{-3}$), we see a systematic divergence
among the $n_{\rm e}$ inferred from different theoretical line profiles.

The calculations we use to determine $n_{\rm e}$ are those of
\citet{Lemke97} \citep[which follow the theory of][VCS]{Vidal73},
\citet[][TB]{Tremblay09}, \citet[][GGC]{Gigosos03}, and Xenomorph
\citep[XENO; see][]{Ferri14,Gomez16,Gomez16b}. These first two are
semi-analytic calculations often used in WD atmosphere
models. These latter two both follow a computer-simulation approach
\citep[e.g.,][]{Stamm84} and have not been used for WD analysis. We do
not include inferences using GGC profiles beyond
$\sim20\times10^{16}$~cm$^{-3}$ because the authors do not claim
validity at values greater than that. We still plot GGC inferences less than
$\sim20\times10^{16}$~cm$^{-3}$ (even though the plotted symbols lie
underneath those using other calculations) because they give credence
to Xenomorph, the other computer-simulation calculation. Note by how
far we exceed the maximum $n_{\rm e}$ achieved by the benchmark
experiment of \citet{Wiese72}, the only other experiment that measured
multiple H Balmer lines near these plasma conditions.

\begin{figure}[!h]
\begin{center}
  \includegraphics[width=\columnwidth]{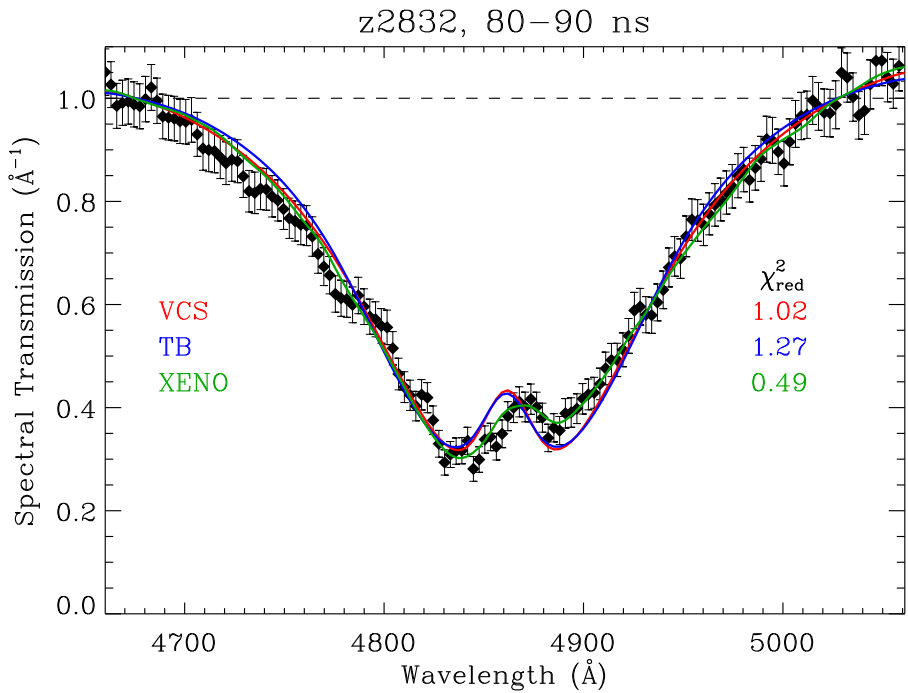}
  \caption{Measured H$\beta$ spectral transmission at 80--90 ns during
  experiment z2832. We fit using different theoretical line-profile
  calculations ($n_{\rm e}\sim83$, $\sim93$, and
  $\sim76\times10^{16}$~cm$^{-3}$ for VCS, TB, and XENO, respectively)
  and show the goodness of fit (reduced $\chi^2$).}
  \label{fit}
\end{center}
\end{figure}

Figure~\ref{fit} shows a fit to our measured H$\beta$ line integrated from
80 to 90~ns during experiment z2832. Here, the spectral line becomes
quite wide because the electron density is so high ($n_{\rm
  e}\sim80\times10^{16}$~cm$^{-3}$). Also, asymmetry in the line
profile---an effect that is rarely considered in WD synthetic spectra
\citep{Halenka15}---becomes apparent. Xenomorph is the only
calculation we use whose profiles are asymmetric because it includes
greater detail when solving for the Coulomb potential of the radiators
in a plasma \citep{Gomez16}. This causes the goodness of fit (reduced
$\chi^2$) to surpass that of VCS and TB. A reduced $\chi^2$ as low as
we show here indicates that we overestimate the noise level, which
determines the uncertainties plotted for each spectral point \citep{Falcon15c}.

\section{Discussion}

The systematic disagreement between our $n_{\rm e}$ inferences using
different line-profile calculations is small at the lower values of
experiment z2553, but it is apparent and greater than the measurement
uncertainties at the higher values of experiment z2832. This is
troubling because we fit the measured H$\beta$ line to diagnose our
plasma conditions. We chose it for two reasons: (1) because its
theoretical line profiles agree with one another at these lower
densities, and (2) because the H$\beta$ spectral line has been
validated by benchmark experiments \citep{Kelleher93}.

While a few benchmark H-line-profile experiments have reached electron
densities greater than $n_{\rm e}=10\times10^{16}$~cm$^{-3}$
\citep[e.g.,][who achieve $n_{\rm e}\sim28$, $\sim16$, and
$\sim14\times10^{16}$~cm$^{-3}$,
respectively]{McLean65,Baessler80,Helbig81}, the highest density
achieved by one that measures multiple lines is that by
\citet{Wiese72}, who reach $n_{\rm e}\sim9\times10^{16}$~cm$^{-3}$;
measuring multiple lines to test relative line shapes and strengths is
a critical requirement for our laboratory investigation of theoretical
line profiles \citep{Falcon15c}. Our experiment now has the
capability of verifying line-profile calculations at these high
electron densities.

\section{Ongoing Work}

As our experiments continue to evolve, our scientific direction
branches out into multiple directions:

\begin{itemize}
\item Absorption lines at high densities are not explicitly apparent
  in most observed WD spectra because they do not exist at the outer
  radii of WD photospheres \citep{Hubeny95}. They are important in the
  integration over the vertical structure of the atmosphere, though,
  which includes a broad range of densities \citep{Hubeny94}, and more
  so for massive WDs \citep[e.g.,][]{Hermes13b}. To determine the
  preferred theoretical line profiles to insert into WD atmosphere
  models, we are now including different ones into atmosphere calculations.

\item Because we spectroscopically observe line profiles in {\it
    absorption}, our measured Balmer lines share the same {\it
    lower}-level population. Using published oscillator strengths
  \citep{Baker08}, this permits us to compare relative line strengths
  between Balmer lines as a way to directly extract occupation
  probabilities. We can then compare our measurements with
  calculations \citep[i.e.,][]{Seaton90}.

\item We now expand our laboratory experiments to other compositions
  relevant to WD photospheres \citep{Falcon13b}. \citet{Schaeuble16}
  show data measured from helium plasmas, which can be used to test
  not only theoretical He line broadening
  \citep[e.g.,][]{Beauchamp97}, but also line shifts relevant to
  gravitational-redshift work \citep{Falcon12}. We also now create
  carbon plasmas whose measured lines can be used to test the
  theoretical line profiles used in atmosphere models for
  carbon-dominated WDs \citep{Dufour11}.
\end{itemize}

\acknowledgements This work was performed at Sandia National
Laboratories. We thank the {\it Z} dynamic-hohlraum, accelerator,
diagnostics, materials-processing, target-fabrication, and wire-array
teams, without which we cannot run our experiments. Sandia is a
multiprogram laboratory operated by Sandia Corporation, a Lockheed
Martin Company, for the United States Department of Energy under
contract DE-AC04-94AL85000. Thank you, P.-E. Tremblay, for providing
the TB theoretical line profiles. T.~A.~G. acknowledges support from the
National Science Foundation Graduate Research Fellowship under grant
DGE-1110007. M.~H.~M. and D.~E.~W. gratefully acknowledge support from
the United States Department of Energy under grant DE-SC0010623 and from
the National Science Foundation under grant AST-1312983. This work has
made use of NASA's Astrophysics Data System Bibliographic Services.


\end{document}